\documentclass[conference]{IEEEtran}
\IEEEoverridecommandlockouts
\usepackage{cite}
\usepackage{amsmath,amssymb,amsfonts}
\usepackage{algorithmic}
\usepackage{graphicx}
\usepackage{textcomp}
\usepackage[table]{xcolor}
\usepackage[hidelinks]{hyperref}
\usepackage{booktabs}
\usepackage{multirow}
\def\BibTeX{{\rm B\kern-.05em{\sc i\kern-.025em b}\kern-.08em
    T\kern-.1667em\lower.7ex\hbox{E}\kern-.125emX}}
\begin{document}
\begin{sloppy}

\title{Stop Words for Processing Software Engineering Documents: Do they Matter?}

\author{
\IEEEauthorblockN{Yaohou Fan}
\IEEEauthorblockA{\textit{The University of Melbourne} \\
Australia \\
yaohouf@student.unimelb.edu.au}
\and
\IEEEauthorblockN{Chetan Arora}
\IEEEauthorblockA{\textit{Monash University}\\
Australia \\
chetan.arora@monash.edu}
\and
\IEEEauthorblockN{Christoph Treude}
\IEEEauthorblockA{\textit{The University of Melbourne} \\
Australia \\
christoph.treude@unimelb.edu.au}
}

\maketitle

\begin{abstract}
Stop words, which are considered non-predictive, are often eliminated in natural language processing tasks. However, the definition of uninformative vocabulary is vague, so most algorithms use general knowledge-based stop lists to remove stop words. There is an ongoing debate among academics about the usefulness of stop word elimination, especially in domain-specific settings. In this work, we investigate the usefulness of stop word removal in a software engineering context. To do this, we replicate and experiment with three software engineering research tools from related work. Additionally, we construct a corpus of software engineering domain-related text from 10,000 Stack Overflow questions and identify 200 domain-specific stop words using traditional information-theoretic methods. Our results show that the use of domain-specific stop words significantly improved the performance of research tools compared to the use of a general stop list and that 17 out of 19 evaluation measures showed better performance.

Online appendix: \url{https://zenodo.org/record/7865748}
\end{abstract}


\section{Introduction}

With the development of natural language processing technology, the amount of text data has exploded in various fields. This phenomenon is also occurring in the field of software engineering, and text analysis using natural language processing (NLP) techniques has become increasingly prevalent in the field of software engineering~\cite{murphy2014function, glier2014exploring}. However, without effective pre-processing methods, raw text corpora often pose computational and analytical barriers to NLP tasks such as indexing, text classification and information retrieval~\cite{kaur2018systematic}. Because of the large number of uninformative words present in the raw text, these words are defined as `stop words'. They appear frequently in text mining tasks and carry little information, simply connecting words in a sentence, such as the words `the', `and', `to', and so on. Theoretically, the removal of stop words can improve the statistical significance of important terms~\cite{manco2002towards, choy2012effective}. Therefore, stop word removal is an important step in text pre-processing. 
 
Using a standard list of stop words to remove these high-frequency and uninformative words has become the standard in research and industry. Researchers often use off-the-shelf generic stop word lists, such as those provided by the Natural Language Processing Toolkit (NLTK), to remove noise~\cite{bird2009natural}. The use of stop word lists is controversial because it is impossible to have a uniform list of stop words and the semantic importance of each word depends on the task and usage. Blind use of stop word lists may result in the loss of important information, which may adversely affect the accuracy of text mining algorithms~\cite{sinka2003towards, makrehchi2008automatic}. However, stop words are difficult to define rigorously, so there are few papers that systematically investigate the impact of stop word removal on algorithm performance. This has led to the current disagreement on whether stop words should be retained or removed. 
 
More reliable lists of stop words might be generated from a specific domain, as common words are distinct in different domains. The technical language used in software engineering texts is different from others; therefore, there might be specific stop words in the texts. Using existing stop lists may result in many useless terms remaining in the data. In fact, experts often maintain an ad hoc stop list~\cite{sarica2020technet, seki2005application} manually. In contrast, few studies have captured the distribution of stop words in a specific domain, i.e., limiting the corpus to a certain domain to extract stop words. 

Inspired by this, we address two research questions (RQs): 
 
\begin{description}
\item[RQ1] Does removing stop words from general stop lists improve the performance of algorithms that have been previously published?
\item[RQ2] Can stop lists generated from the software engineering (SE) domain positively affect downstream tasks in their domain?
\end{description}
 
To answer RQ1, this study replicated the tools proposed in three previous papers in the SE domain. After this, stop lists of different lengths were used to pre-process the raw text data and reproduce all tools. The analysis of the impact on the output of the tools was completed by looking at the proportion of words removed in the text data and measures of the performance of the tools used in the original papers. 
 
We explored the SE-related corpus and used two automated methods to statistically count stop words to answer RQ2. These domain-specific stop words are examined in the same way in the replicated tools. We found that using stop lists generated from the public domain had a positive effect on the algorithms' output in some evaluation metrics, but a negative effect in others. SE domain-specific stop word lists performed better than general lists, but it is important to take context into account before removing stop words as blind use can have a negative impact. We conclude the paper with a systematic comparison and analysis of all stop lists. All stop word lists and scripts are available in our online appendix: \url{https://zenodo.org/record/7865748}

\section{Related Work}

Research on stop words dates back to 1957, when H.P. Luhn first introduced the concept of stop words~\cite{luhn1957statistical}. Subsequent studies have systematically studied stop words and summarised seven properties~\cite{kaur2018stopwords}. Stop words are generally a single set of words, and most stop lists contain definite articles and coordinating conjunctions. Researchers have worked to extract stop words from general knowledge bases. The Brown Corpus is the most widely used corpus of general language, which contains about one million words. Many stop lists for general language have been generated from this corpus~\cite{fox1989stop, maverick1969computational}. In addition, Montemurro and Zanette~\cite{montemurro2010towards} learned linguistic structures in a large corpus of written languages and exported a stop list. 
 
Static stop lists drawn from a public knowledge base are seen by many practitioners as the standard and the default choice. However, Lo et al.~\cite{lo2005automatically} pointed out that `standard' stop lists ignore domain knowledge properties. These `standard' stop lists have become outdated over time, with the first English stop lists published in the 1970s~\cite{van1979information}. The usage of some popular expressions has changed, due to social factors such as technological and cultural shifts. It is vital to update and revise the stop lists for current use. Other studies have similarly identified the drawbacks of `standard' stop words~\cite{chakrabarti1997using, silva2003importance}. Experts have proposed many heuristics to overcome these limitations. The methods commonly used in the industry can be divided into the two categories discussed in the following.
 
\subsection{Methods based on Zipf's Law}
 
The study of word frequency in texts first appeared in the last century when George K.~Zipf customised the rule of thumb to rank lists of words using the frequency of terms (tf) as a measure~\cite{zipf1949selective}. A study borrowed this concept by statistically calculating the relative importance of individual words and phrases~\cite{luhn1957statistical}, further developing the hypothesis that the most discriminatory words should appear in the middle of the frequency rank. Based on these frequency descriptions, Zipf's law approach can be summarised as: 

\begin{itemize}
\item Term-frequency High (TF-high): remove the most frequently occurring words.  
\item Term-frequency 1 (TF1): remove words that occur once.  
\item Inverse document frequency (IDF): remove words that occur frequently in multiple documents. 
\end{itemize}
 
Myerson~\cite{myerson2013fundamentals} combined TF-high and statistical methods to obtain 500 stop words from a Chinese corpus. The first step of their method was to filter the list of candidate stop words using the TF-high method. After this, the statistical correlation (Chi-squared statistic) between the candidate words and the text categories was calculated, and words with low correlation were identified as stop words. Their results showed that 500 stop words effectively removed 43\% of the words from the corpus and improved the accuracy of the classification task by 7\%. Salton and Buckley~\cite{salton1988term} combined the TF-high and IDF methods to propose a method called TF-IDF, which is effective in representing the weight of terms in an arbitrary setting, where the weight is also the usefulness. 
 
\subsection{Machine learning-based methods}
 
The main objective of machine learning methods is to reduce the dimensionality of the document vector representation in the feature space~\cite{aizawa2003information}. They use a method of ranking terms to eliminate words that do not contain information. All term ranking methods are based on three components: (a) calculating importance values for each word based on statistical indicators; (b) ranking features in decreasing order of importance; and (c) establishing a threshold to filter the ranked list. Commonly used statistical indicators are: 

\begin{itemize}
\item Information gain (IG)~\cite{yang1999re}: IG is the difference between the information entropy of a feature (word) before and after it appears in the text, and it considers the information load of the feature in the text. The smaller the load, the less important the feature becomes.  
\item Odds ratio~\cite{mladenic1998feature}: It is a statistical measure that quantifies the strength of the association between two events (word and text). When the presence or absence of a word does not change the semantics of the text, they are independent of each other, so the word can be classified as a stop word. 
\item Expected cross entropy~\cite{koller1997hierarchically}: Cross entropy measures the information about the variability between two probability distributions, and in the calculation of stop lists, it indicates how difficult a word is for text classification. Therefore, words that do not contribute to text classification are evidently stop words. 
\end{itemize}
 
Asubiaro~\cite{asubiaro2013entropy} proposed an entropy-based method to determine the probability that Yoruba words are stop words. Their experiment calculated entropy for all words with non-noun lexical forms, and those with values greater than 0.6 were identified as stop words. The results show that a stop list of length 256 removes 66\% of the words in the corpus, and they indicated that after translating the list into English, only 69.1\% of the stop words appeared in the common English stop list. 

\section{Replicated Tools and Their Evaluation Metrics}

This paper replicates three tools implemented using NLP techniques relevant to the field of software engineering: an automated app review classifier~\cite{maalej2016automatic}, a natural language query recommendation tool~\cite{rahman2017rack}, and a requirements change impact analysis tool~\cite{arora2015change}, with the aim of answering RQ1 and RQ2 using different stop lists in these three downstream tasks. We selected these tools based on the availability and operability of their replication packages.

\subsection{Tool 1 - App Reviews Classification}

Numerous studies have investigated the classification of texts related to software engineering, such as Application Programming Interface (API) documents~\cite{petrosyan2015discovering} and README files~\cite{prana2019categorizing}. Maalej et al.~used machine learning algorithms to build classifiers of user reviews, thus helping developers filter and process useful information from reviews~\cite{maalej2016automatic}. 

\begin{itemize}
\item Data: (1) Their experimental data consists of 4400 reviews from the Apple App Store and Google Play Store. These reviews were divided into four categories, 737 User Experience (UE), 378 Bug Report (BP), 299 Feature Request (FR), and 2721 Rating (RT). (2) List of stop words used to purify text: NLTK stop words (179 words). 
\item Model: A Naïve Bayes binary classifier was constructed for each of the four categories, and the results were then aggregated to obtain the final classification output. 
\item Evaluation Metrics:
\begin{itemize}
\item Precision: refers to the proportion of samples classified as positive that are actually positive. 
\item Recall: is the fraction of the true positive sample that was retrieved. 
\item F1 measure: Harmonised mean of precision and recall. 
\end{itemize}
\end{itemize}

Note that these reproduced models did not perform exactly the same as reported in previous work~\cite{maalej2016automatic}, since we were unable to access their source code. We refer readers to Maalej et al.~\cite{maalej2016automatic} for the comparison. Only the replicated model from our own implementation was used for subsequent analysis and evaluation. 

\subsection{Tool 2 - Query Recommendation Tool (RACK)}

API retrieval is another application of NLP techniques in the field of software engineering. API retrieval tools accept natural language queries from developers as input and then return a set of related API classes or methods automatically. A query recommendation tool, called RACK, proposed by Rahman and Lo~\cite{rahman2017rack}, first captures the intention of queries from code annotations in the IDE and then converts these queries into relevant API classes. 

\begin{itemize}
\item Model: Such a conversion is achieved in two steps with the Keyword-API association in Stack Overflow and the Query Reformulation:  
\begin{itemize}
\item Keyword-API Mapping Database: 344K pairs of questions and answers from the Stack Overflow website were analysed to generate intrinsic associations between the keywords in question and the API classes, and the associations were used to build the Keyword-API Mapping Database. 
\item Query Reformulation: In this step, the raw unstructured natural language query input is pre-processed (i.e., tokenization, stop word removal, stemming) using NLP techniques, and a keyword vector is obtained; RACK accesses the database to collect candidate API classes and sorts them heuristically to output a sorted list of relevant API classes. 
\end{itemize}
\item Data: The authors provide 175 natural language queries with ground truth. This study used these queries, then evaluated the accuracy of the recommendations by varying the stop list used in the pre-processing phase. In addition, a stop list of length 797 is used to reformulate the queries. 
\item Evaluation Metric: 
\begin{itemize}
\item Top-10: Refers to the percentage of search queries in which at least one API class was correctly recommended by the recommendation technology in the Top-10 results. 
\item Mean Reciprocal rank@10 (MRR@10): Reciprocal Rank@10 is the multiplicative reciprocal of the first relevant API class ranking in the Top-10 results returned by the tool. Mean Reciprocal Ranking@10 (MRR@10) is the average of such metrics across all search queries in the dataset. 
\item Mean Average Precision@10 (MAP@10): Precision@10 is the precision of all relevant APIs that appear in the Top-10 recommendation list. Average Precision@10 (AP@10) is the average Precision@10 for relevant APIs recommended for a natural language (NL) query. Mean Average Precision@10 is the average of the AP@10 for all NL queries. 
\item Mean Recall@10 (MR@10): Recall@10 represents the percentage of Top-10 results for which this tool correctly recommends all relevant APIs for an NL query. Mean Recall@10 (MR@10) is the average of these measures for all queries. 
\end{itemize}
\end{itemize}

\subsection{Tool 3 - Requirements Change Impact Analysis}

During the development of software, requirements are frequently changed to ensure that the final product meets expectations. Changing one requirement in a requirements specification requires the entire specification to be modified to keep it correct and consistent. Manually analysing the impact of a change to one requirement on other requirements is time consuming. A natural language processing approach has been proposed to deal with the impact of requirement changes~\cite{arora2015change}, which takes into account the phrase structure of requirement statements and captures the change propagation conditions to achieve a more accurate analysis of the impact of changes. The proposed approach is given by Arora et al.~\cite{arora2015change,arora2015narcia}.

\begin{itemize}
\item Data: The original paper tested the proposed approach using 14 scenarios from the requirements specifications of two industrial cases (case A and case B). For this study, we only tested three scenarios from case B due to resource limitations: query 2, query 4, and query 5. 
\item Stop list: Generated from Brown Corpus 
\item Evaluation metrics: Area under the curve (AUC): Commonly used in statistics and machine learning to assess the performance of a binary classification model by measuring the ranking performance of the model. Ranking performance refers to the probability of ranking a positive sample ahead of a negative sample in classification. In this experiment, it assesses the quality of the ranking of the impacted requirements. 
\end{itemize}

\section{Stop Word List Construction}

To answer RQ2, we used a corpus from Stack Overflow, a programming question and answer website. Today, the field of software engineering is made up of a wide variety of technologies, languages, and platforms, and developers are required to have a broad skill set. Even experienced senior software engineers find it difficult to keep up with the pace of evolution of technology. Thus, Stack Overflow leverages the expertise of its users to provide answers to technical concerns, and over time, it has become a repository of software engineering knowledge. Meanwhile, the question-and-answer text on the site is generated by engineers, and much of the text consists of technical terms. It is therefore one of the largest corpora in software engineering. Some studies have mined the textual information contained therein, e.g.,~\cite{treude2016augmenting}. 
 
We crawled the Stack Overflow question API with four popular tags (`javascript', `css', `pythonv2' and `deep learning'), focusing on questions that have been answered. We sort these questions by the number of times they have been viewed, taking the top 2500 from each category, for a total of 10,000 questions crawled as a corpus for the software engineering domain. 

\subsection{Data processing}

So far, the natural language text in the corpus is unstructured and cleaning the original text using regular expressions is necessary to improve the accuracy of stop word identification. The most common pre-processing steps are applied: (1) removing mentions, (2) removing URLs, (3) removing hashtags, (4) removing numbers, (5) removing text in brackets, (6) removing non-English words, (7) removing extraneous space. 
 
After this, the size of the corpus is reduced by lowercasing and lemmatizing all words. For grammatical reasons, the original corpus contains different forms of a word, such as see, saw, and seen, but they have the same semantic meaning. Lemmatization analyses the morphology of the word, thus removing the flexion endings to return the basic form of the word. 

\subsection{TF-IDF based method for identifying stop words}

There is a long history of using the TF-IDF method and its variants to separate stop words and content words from the corpus. Extensive experience has shown it to be an effective method~\cite{fox1989stop, aizawa2000feature, konchady2006text}. TF-IDF is the product of two quantities TF and IDF. Based on information theory, TF estimates the probability of occurrence of a term when normalised by the total frequency in the document or set of documents. The calculation of term frequency of the i-th term in the j-th document is given by: 

\begin{equation}
tf_{ij}=log(1+freq(t,d))
\end{equation}

The inverse document frequency is calculated by giving the ratio of the total number of documents $|D|$ to the amount of documents that contain the term $t$ in logarithmic form, and hence the importance $t$ in the document is denoted by the following expression: 

\begin{equation}
idf=log(\frac{|D|}{count(d\in D:t\in d)})
\end{equation}

The importance of a word is expressed as the product of the probability of its occurrence and the amount of information it represents. This paper identifies words that occur in many documents (high df) and words that occur less frequently (low tf) as stop words and selects the top 200 words for the stop list based on TF-IDF scores. 

\subsection{Stop words recognition based on Poisson distribution}

On the other hand, Church and Gale~\cite{church1995poisson} have studied the statistical distribution of words and have proposed the hypothesis that stop words follow a Poisson distribution. We refer to this statistical method of computing stop word lists as a Poisson stop list. There have been studies that validate this unsupervised method by creating a stop list in a corpus of Polish~\cite{jungiewicz2014unsupervised}. Under their assumptions, the document frequency ($df$) of words can be approximated from the term frequency ($tf$) and the number of documents ($N$), by using probability theory: 

\begin{equation}
\frac{Estimated\ df}{N}=1-P(term\ does\ not\ occur)
\end{equation}

Assuming a Poisson distribution for stop words, with $\mu$ as the average number of occurrences of a word in each document, the probability of the word occurring $k$ times is:  

\begin{equation}
\mu=\frac{tf}{N}
\end{equation}

\begin{equation}
P(k,\mu)=\frac{e^{-\mu}*\mu^k}{k!}
\end{equation}

Randomly distributed words (stop words) should yield an estimated $df/df$ close to 1, while highly cluttered words (key words) show an increasing trend in value. In this paper, the words in the corpus with an estimated $df/df$ of 1 were calculated as candidate stop words, and after this a stop list of 200 length was randomly selected in order to keep the lengths of the stop lists consistent.
 
Figure~\ref{fig:flowchart} illustrates the process of generating a stop list from an SE domain-specific corpus for this paper. First, the raw data is pre-processed and then the lists are ranked using statistical metrics (TF-IDF and Poisson distribution). Following this, the first author leveraged his domain expertise to manually eliminate unstructured terms, as the regular expressions did not fully clean the text.
Finally, a final list related to the software engineering domain is completed. 

\begin{figure}
    \centering
    \includegraphics[scale=0.7]{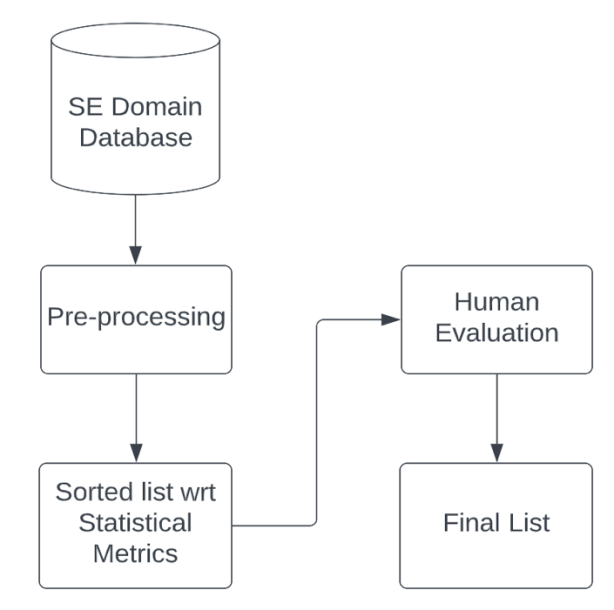}
    \caption{The flowchart for generating a domain-specific stop list}
    \label{fig:flowchart}
\end{figure}

\subsection{Other stop word lists used in the experiment}

To answer RQ1, we collected five existing stop lists that are publicly available, as shown in Table~\ref{table:one}, four of which are generated from public knowledge bases and widely used in NLP pre-processing tasks. The last stop list is a stop list in a technical domain~\cite{sarica2021stopwords}. 

\begin{table*}[]
\centering
\caption{Other freely available stop word lists used in this experiment}
\begin{tabular}{llrlll}
\toprule
ID & Stop word list & Length & & Source URL \\
\midrule
1  & Ranks NL (Google)                                              & 32 & (very small)                                                  & http://www.ranks.nl/stopwords                                                                                 \\
2  & MongoDB                                                        & 174 & (small)                                                     & https://github.com/mongodb/mongo/blo b/master/src/mongo/db/fts/stop\_words\_ english.txt                      \\
3  & Azure Gallery                                                  & 311 & (medium)                                                     & https://gallery.azure.ai/Experiment/Ho w-to-modify-default-stopword-list-1                                    \\
4  & Alir3z4                                                        & 1298 & (large)                                                      & https://github.com/Alir3z4/stopwords/blob/master/english.txt \\
5  & Technology Domain                                              & 87   &                                                              & https://journals.plos.org/plosone/article ?id=10.1371/journal.pone.0254937                                    \\
\bottomrule
\label{table:one}
\end{tabular}
\end{table*}

Stop word lists 1-4 are not related to a specific domain; they are generated from language in common use and range in length from 32 to 1298. We sort the lengths of stop word lists from small to large and define them as `very small', `small', `medium', and `large'. We note that there is no existing research to divide the lengths of stop word lists. The reason for doing this is to facilitate subsequent comparisons. 
 
Sarica and Luo~\cite{sarica2021stopwords} collected natural language texts that describe technologies at all levels to identify stop words in technical language texts and constructed a database. They combined algorithms such as TF, TF-IDF, and information entropy to automatically identify candidate stop words. Human experts then evaluated the importance of these candidates in filtering out the real stop words. 

\subsection{Stop word Removal}

We used the classic stop word removal method to sanitise the raw data. The classic algorithm accepts an English text as input and outputs the text after all stop words have been removed. The process is illustrated in Figure~\ref{fig:workflow}. 

\begin{figure}
    \centering
    \includegraphics[width=\linewidth]{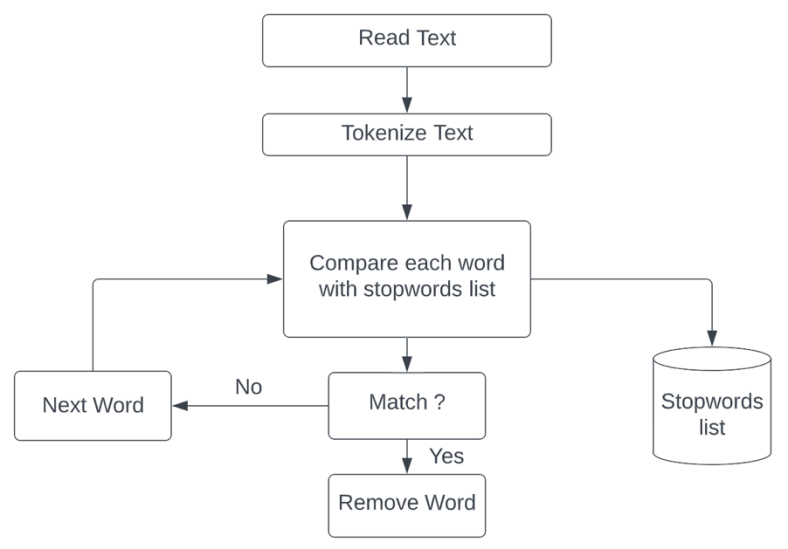}
    \caption{Classic workflow of stop word removal}
    \label{fig:workflow}
\end{figure}

\section{Experiment Analysis}

\subsection{Tool 1 - App Reviews Classification}

\subsubsection{Stop list Evaluation}

Eliminating stop words usually results in a significant reduction in the number of words in the corpus, allowing for a more accurate approach to text mining. This paper summarises how many stop words were removed from the raw text data using different stop lists for this replicated tool. Note that after lowercasing and lemmatizing of the raw text data, there are 9655 unique words remaining. 

\begin{table}[]
\centering
\caption{Number of stop words removed for app review classification tool}
\begin{tabular}{lrrr}
\toprule
Stop list              & Length & words remaining & \% removed \\
\midrule
Very small            & 32     & 7849                                                                                                   & 18.70\%                           \\
Small                 & 174    & 7734                                                                                                   & 19.89\%                           \\
Medium                & 311    & 7656                                                                                                   & 20.70\%                           \\
Large                 & 1298   & 7257                                                                                                   & 24.84\%                           \\
Technology   Domain   & 87     & 7823                                                                                                   & 18.97\%                           \\
SE Domain   (TF-IDF)  & 200    & 7617                                                                                                   & 21.10\%                           \\
SE Domain   (Poisson) & 200    & 7612                                                                                                   & 21.16\%                           \\
\bottomrule
\label{table:two}
\end{tabular}
\end{table}

Table~\ref{table:two} shows that the number of words retained after pre-processing is negatively correlated with the length of the stop list, and a large stop list significantly removes more stop words than a small stop list. Additionally, the stop lists generated in the SE domain resulted in fewer words remaining compared to the longer stop lists (medium), which implicitly indicates that they contain more accurate domain-related stop words. 

\subsubsection{Model Performance Evaluation}

\begin{table*}[]
\centering
\caption{Performance of general stop list for app review classification tool. Green indicates better performance than the original, orange worse performance than the original.}
\begin{tabular}{l|rrr|rrr|rrr|rrr}
\toprule
\multirow{2}{*}{}                                                                  & \multicolumn{3}{c}{PD (bug report)} & \multicolumn{2}{c}{RT (rating)} & \multicolumn{3}{c}{FR (feature request)} & \multicolumn{3}{c}{UE (user experience)} \\
                                                                                   & Pre        & Rec        & F1        & Pre            & Rec            & F1     & Pre         & Rec          & F1          & Pre          & Rec         & F1          \\
\midrule
Original (NLTK stop list) & 9.9\%      & 29.5\%     & 14.8\%    & 73.8\%         & 72.9\%         & 73.3\% & 5.7\%       & 17.9\%       & 8.7\%       & 11.6\%       & 27.4\%      & 16.3\%      \\
No stop list & \cellcolor{green}10.8\%     & \cellcolor{orange}28.6\%     & \cellcolor{green}15.7\%    & \cellcolor{green}75.6\%         & \cellcolor{orange}71.0\%         & 73.3\% & \cellcolor{orange}5.3\%       & \cellcolor{orange}15.5\%       & \cellcolor{orange}7.9\%       & \cellcolor{orange}8.7\%        & \cellcolor{orange}18.8\%      & \cellcolor{orange}11.9\%      \\
Very small                                                                         & \cellcolor{green}12.2\%     & \cellcolor{green}35.7\%     & \cellcolor{green}18.2\%    & \cellcolor{green}74.9\%         & \cellcolor{green}74.8\%         & \cellcolor{green}74.9\% & 5.7\%       & \cellcolor{green}19.0\%       & \cellcolor{green}8.8\%       & \cellcolor{orange}10.3\%       & \cellcolor{orange}23.9\%      & \cellcolor{orange}14.4\%      \\
Small                                                                              & \cellcolor{green}12.7\%     & \cellcolor{green}50.0\%     & \cellcolor{green}20.3\%    & \cellcolor{orange}72.1\%         & \cellcolor{green}78.2\%         & \cellcolor{green}75.0\% & \cellcolor{green}6.8\%       & \cellcolor{green}28.6\%       & \cellcolor{green}10.9\%      & \cellcolor{green}12.2\%       & \cellcolor{green}34.0\%      & \cellcolor{green}18.0\%      \\
Medium                                                                             & \cellcolor{green}12.7\%     & \cellcolor{green}52.7\%     & \cellcolor{green}20.5\%    & \cellcolor{orange}71.0\%         & \cellcolor{green}81.0\%         & \cellcolor{green}75.6\% & \cellcolor{green}7.8\%       & \cellcolor{green}34.5\%       & \cellcolor{green}12.7\%      & \cellcolor{orange}11.2\%       & \cellcolor{green}32.0\%      & \cellcolor{green}16.6\%      \\
Large                                                                              & \cellcolor{green}10.4\%     & \cellcolor{green}56.2\%     & \cellcolor{green}17.5\%    & \cellcolor{orange}68.6\%         & \cellcolor{green}83.1\%         & \cellcolor{green}75.1\% & \cellcolor{green}7.5\%       & \cellcolor{green}48.8\%       & \cellcolor{green}13.0\%      & \cellcolor{green}12.5\%       & \cellcolor{green}45.2\%      & \cellcolor{green}19.6\%  \\
\bottomrule
\label{table:three}
\end{tabular}
\end{table*}

\begin{table*}[]
\centering
\caption{Performance of the domain-specific stop list for the app review classification tool}
\begin{tabular}{l|rrr|rrr|rrr|rrr}
\toprule
\multirow{2}{*}{}                                                      & \multicolumn{3}{c}{PD (bug report)} & \multicolumn{3}{c}{RT (rating)}  & \multicolumn{3}{c}{FR (feature request)} & \multicolumn{3}{c}{UE (user experience)} \\
                                                                       & Pre        & Rec        & F1        & Pre            & Rec            & F1     & Pre         & Rec          & F1          & Pre          & Rec         & F1          \\
\midrule
Original                                                               & 9.9\%      & 29.5\%     & 14.8\%    & 73.8\%         & 72.9\%         & 73.3\% & 5.7\%       & 17.9\%       & 8.7\%       & 11.6\%       & 27.4\%      & 16.3\%      \\
Technology domain                                                      & \cellcolor{green}10.5\%     & \cellcolor{green}30.4\%     & \cellcolor{green}15.6\%    & \cellcolor{green}75.4\%         & \cellcolor{orange}71.2\%         & \cellcolor{orange}73.2\% & \cellcolor{orange}4.4\%       & \cellcolor{orange}14.3\%       & \cellcolor{orange}6.7\%       & \cellcolor{green}12.2\%       & \cellcolor{green}34.0\%      & \cellcolor{green}18.0\%      \\
SE domain (Poisson) & \cellcolor{green}10.0\%     & \cellcolor{green}37.5\%     & \cellcolor{green}15.8\%    & \cellcolor{orange}72.1\%         & \cellcolor{green}78.0\%         & \cellcolor{green}74.9\% & \cellcolor{green}7.1\%       & \cellcolor{green}29.8\%       & \cellcolor{green}11.5\%      & 11.6\%       & \cellcolor{green}32.0\%      & \cellcolor{green}17.0\%      \\
SE domain (TF-IDF) & \cellcolor{green}10.7\%     & \cellcolor{green}40.2\%     & \cellcolor{green}16.9\%    & \cellcolor{orange}72.2\%         & \cellcolor{green}78.2\%         & \cellcolor{green}75.1\% & \cellcolor{green}7.9\%       & \cellcolor{green}33.3\%       & \cellcolor{green}12.8\%      & \cellcolor{green}11.7\%       & \cellcolor{green}32.5\%      & \cellcolor{green}17.2\%      \\
\bottomrule
\label{table:four}
\end{tabular}
\end{table*}

Tables~\ref{table:three} and~\ref{table:four} represent the performance of the models obtained after pre-processing with different stop lists, respectively. The small and large stop lists are the best general lists for Tool 1, as their models perform better overall, outperforming the replicated model on 11 of 12 measures. However, it is interesting to note that as the length of the general stop list grows, the precision metric in the RT classification tends to decrease, when all other metrics are getting better. The opposite trend is observed in the Recall indicator, where recall values increase with the length of the stop list in almost all four categories. 
 
It is also clear that stop words extracted from the SE domain perform better than those from the technical domain, which is to be expected since there is SE-related text in user application reviews; for example, some SE domain lexicons appear in feature request reviews. Furthermore, domain-specific stop lists performed no better than generic ones of the same volume in this downstream task. This is because most of the vocabulary in the user review text data is derived from the knowledge base of the end user, meaning that the body of the text is still composed of general language. 

\subsection{Tool 2 - Query Recommendation Tool (RACK)}

\subsubsection{Stop list Evaluation}

\begin{table}[]
\centering
\caption{Number of stop words removed for query recommendation tool}
\begin{tabular}{lrrr}
\toprule
Stop list              & Length & words remaining  & \% removed \\
\midrule
Very small            & 32     & 389                                                                                                    & 4.18\%                            \\
Small                 & 174    & 380                                                                                                    & 6.40\%                            \\
Medium                & 311    & 369                                                                                                    & 9.85\%                            \\
Large                 & 1298   & 342                                                                                                    & 15.76\%                           \\
Technology   Domain   & 87     & 400                                                                                                    & 1.48\%                            \\
SE Domain   (TF-IDF)  & 200    & 328                                                                                                    & 19.21\%                           \\
SE Domain   (Poisson) & 200    & 326                                                                                                    & 20.44\%                           \\
\bottomrule
\label{table:five}
\end{tabular}
\end{table}

Table~\ref{table:five} shows the status of the lexicon in the corpus after the removal of stop words using different stop lists. For the API recommendation tool, stop lists identified from the general knowledge base perform poorly, and the longest stop list removes only 15.76\% of the words. In contrast, the SE domain-specific stop list with only 200 words removed about 20\% of the words in the original text. The stop list generated from the technology domain does not contain relevant stop words, as it removed almost no words from the text. 

\subsubsection{Model Performance Evaluation}

\begin{table}[]
\centering
\caption{Performance of the general stop list for query recommendation tool}
\begin{tabular}{lrrrr}
\toprule
             & Top-10  & MRR@10  & MAP@10  & MR@K    \\
\midrule
Original     & 83.43\% & 52.29\% & 45.73\% & 54.07\% \\
No stop words& \cellcolor{orange}81.14\% & \cellcolor{orange}50.72\% & \cellcolor{orange}44.94\% & \cellcolor{orange}51.10\% \\
Very Small   & \cellcolor{orange}81.14\% & \cellcolor{orange}50.72\% & \cellcolor{orange}44.94\% & \cellcolor{orange}51.10\% \\
Small        & \cellcolor{orange}81.14\% & \cellcolor{orange}50.72\% & \cellcolor{orange}44.94\% & \cellcolor{orange}51.10\% \\
Medium       & \cellcolor{orange}81.14\% & \cellcolor{orange}50.97\% & \cellcolor{orange}44.83\% & \cellcolor{orange}51.91\% \\
Large        & \cellcolor{orange}80.57\% & \cellcolor{orange}49.71\% & \cellcolor{orange}44.07\% & \cellcolor{orange}51.03\% \\
\bottomrule
\label{table:six}
\end{tabular}
\end{table}

\begin{table}[]
\centering
\caption{Performance of domain-specific stop list for query recommendation tool}
\begin{tabular}{lrrrr}
\toprule
                      & Top-10  & MRR@10  & MAP@10  & MR@K    \\
\midrule
Original              & 83.43\% & 52.29\% & 45.73\% & 54.07\% \\
Technology domain     & \cellcolor{orange}81.14\% & \cellcolor{orange}50.72\% & \cellcolor{orange}44.94\% & \cellcolor{orange}51.10\% \\
SE domain  (Poisson) & \cellcolor{green}83.85\% & 52.29\% & \cellcolor{orange}43.27\% & \cellcolor{green}54.47\% \\
SE domain (TF-IDF)    & \cellcolor{green}84.17\% & \cellcolor{green}53.20\% & \cellcolor{green}45.82\% & \cellcolor{green}56.18\% \\
\bottomrule
\label{table:seven}
\end{tabular}
\end{table}

From the experimental results (Table~\ref{table:six} and Table~\ref{table:seven}), the size of the stop word list has very little effect on the performance of the model, with all general stop lists giving similar performance. The model even performs stable with no stop word removal. However, other stop lists, such as the large stop list, have a negative impact on the model behaviour. We speculate that the large stop list removed useful information from the data, resulting in a weaker performance of the model. In addition, the smaller stop lists (very small and small) gave the same results as those without the stop list. This suggests that, in the context of this task, these lists did not contain the appropriate stop words and did not reformulate the query well. This validates our previous hypothesis that the definitions of stop words are different between domains and that stop words generated from generic knowledge bases would not maximize the improvement of the algorithm in domain-specific tasks. 
 
The stop list from the SE domain was superior to any other list, and the TF-IDF method outperformed the original performance on all four metrics. Furthermore, the Poisson stop list exceeded the original performance only for the Top-10 and MR@K metrics. On the contrary, the model obtained using the stop list of the technology domain performed the same as if the stop words were not removed, indicating that this stop list does not contain the appropriate stop words of the SE domain. 

\subsection{Tool 3 - Requirements Change Impact Analysis}

\begin{table}[]
\centering
\caption{Performance (normalized AUC) of the general stop list for Tool 3}
\begin{tabular}{lrrrrrr}
\toprule
        & Original & None & V.~Small & Small & Medium & Large \\
\midrule
Query 2 & 0.585    & 0.585        & \cellcolor{green}0.588      & \cellcolor{green}0.588 & \cellcolor{green}0.595  & \cellcolor{orange}0.538 \\
Query 4 & 0.981    & 0.981        & 0.981      & 0.981 & 0.981  & \cellcolor{orange}0.961 \\
Query 5 & 0.561    & \cellcolor{green}0.579        & \cellcolor{green}0.579      & \cellcolor{green}0.577 & \cellcolor{orange}0.447  & \cellcolor{orange}0.462 \\
\bottomrule
\label{table:eight}
\end{tabular}
\end{table}

\begin{table}[]
\centering
\caption{Performance (normalized AUC) of the domain-specific stop list for Tool 3}
\begin{tabular}{lrrrr}
\toprule
    & Original & Technology domain & SE domain & SE domain \\
    & & & (TF-IDF) & (Poisson) \\
\midrule
Query 2 & 0.585    & \cellcolor{green}0.588             & \cellcolor{green}0.588              & \cellcolor{green}0.588                 \\
Query 4 & 0.981    & 0.981             & 0.981              & 0.981                 \\
Query 5 & 0.561    & \cellcolor{green}0.579             & \cellcolor{green}0.602              & \cellcolor{green}0.602                 \\
\bottomrule
\label{table:nine}
\end{tabular}
\end{table}

Table~\ref{table:eight} shows a very clear trend in the experimental outcomes, with all of the common stop word lists performing very similarly, except for the medium stop list and the large one. The medium stop list appeared to be weak in Query 5, only obtaining an AUC of 0.447 down from the original performance of 0.561. However, it performed the best of all stop lists in Query 2, achieving an AUC close to 0.6. The large stop list performed poorly in all three scenarios, which is consistent with the experimental finding for Tool 2, where some larger stop lists removed some valuable words. Furthermore, we found that the model performed surprisingly better in Query 5 without stop word removal.  
 
Domain-specific stop lists improved the performance of the model (Table~\ref{table:nine}), and all lists outperformed the original model in two of the three cases. SE domain-specific stop words had the most significant effect on the model performance, producing an AUC value of 0.602 in Query 5 compared to the original performance of 0.561. 

\subsection{Overall Stop list Performance Comparison}

\begin{table}[]
\centering
\caption{Overall performance}
\begin{tabular}{lrrr}
\toprule
Stop word list & better & worse & same \\
\bottomrule
No stop words                                                        & \cellcolor{green}4                                     & \cellcolor{orange}12                                   & 3                                   \\
Very Small                                                           & \cellcolor{green}10                                    & \cellcolor{orange}7                                    & 2                                   \\
Small                                                                & \cellcolor{green}13                                    & \cellcolor{orange}5                                    & 1                                   \\
Medium                                                               & \cellcolor{green}11                                    & \cellcolor{orange}7                                    & 1                                   \\
Large                                                                & \cellcolor{green}11                                    & \cellcolor{orange}8                                    & 0                                   \\
Technology domain                                                     & \cellcolor{green}9                                     & \cellcolor{orange}9                                    & 1                                   \\
SE domain (Poisson) & \cellcolor{green}12                                    & \cellcolor{orange}5                                    & 2                                   \\
SE domain (TF-IDF) & \cellcolor{green}17                                    & \cellcolor{orange}1                                    & 1    \\
\bottomrule
\label{table:ten}
\end{tabular}
\end{table}

The overall performance of the stop lists generated based on a total of 19 metrics (i.e., all evaluation metrics used by the three replicated tools) used in the experiment is given in Table~\ref{table:ten}. The fact that 12 of the metrics obtained worse results without the stop list is an indication that the stop list can have a positive effect on the model output. We see that among the non-SE domain-specific stop lists, the small stop list with 174 stop words gives the best performance, improving the performance of the tools on 13 metrics. The other stop lists have less of a positive effect on the performance of the tools, as some indicators show worse performance. The stop lists generated from the corpus of other domains (technology domain) performed negatively. Finally, the SE domain-related stop lists gave the best performance, especially the one generated using TF-IDF, which had a positive impact on the model for 17 of the 19 metrics. 

\section{Conclusion}
This paper evaluates the performance of existing tools by removing stop words in three downstream tasks related to the SE domain using various stop lists. Combining the experimental results and analysis, we can provide answers to our two Research Questions in this section. 
 
The experimental results in Table~\ref{table:ten} answer RQ1, whether general stop lists have a positive effect on the performance of algorithms' evaluation measures and tasks. While using stop lists generated from the public domain had a positive effect on the algorithms' output in some of the evaluation metrics, other metrics showed that the model gave worse results. For example, a larger stop list increased the F1 measure value for Tool 1 but led to a decrease in recall. On the other hand, experiments conducted in Tool 2 and Tool 3 showed that all `standard' stop lists had little effect on the algorithm and the model performed similarly to that without the removal of stop words. For RQ2, the SE domain-related stop list showed a positive effect for all three tools and led to the best overall performance. Stop lists generated using TF-IDF outperformed the Poisson method, with 17 of 19 indicators performing better than the original. Although the Poisson method performs similarly to the small stop list, it is superior to other general lists. Therefore, we argue that SE domain-specific stop word lists performed better than general lists. 
 
An overall conclusion can be drawn that it is not possible to have a standard list of stop words. The semantics of each word are different in each domain. The context of the task needs to be taken into account before removing stop words, and the use of domain-specific stop lists can have a positive effect. However, blind use of a stop list may have a negative impact on the results of the algorithm. 

\section{Limitations and Future Improvement}

Future studies can contribute in two directions, domain-specific corpora and algorithms for identifying stop words. The text in the domain-specific corpus used for this paper comes from Stack Overflow. However, such a corpus is biased because there is a wide variety of programming languages used in SE. Therefore, a corpus with more data and broader coverage specific to the SE domain needs to be constructed to support more accurate stop word generation. 
 
On the other hand, we did not use more advanced algorithms to generate stop words to exclude variants, and only the most commonly used methods were employed. These traditional methods are flawed because they are based on some naive assumptions that do not hold in all cases. The current experiment has demonstrated the positive effect of domain-specific stop words in SE using a baseline approach. Future work could investigate stop words in their domain further using more advanced algorithms as well as the impact of different stop word lists on different software engineering tasks.

\section{Data Availability}

All stop word lists and scripts are available in our online appendix: \url{https://zenodo.org/record/7865748}


\end{sloppy}
\end{document}